\documentclass[twocolumn,showpacs,prx,preprintnumbers,amsmath,amssymb,floatfix,eqsecnum]{revtex4}
\usepackage{graphicx}
\usepackage{epsfig,psfrag}
\usepackage{dcolumn}
\usepackage{bm}
\usepackage{color}
\begin{document}
\title{ Orbital ferromagnetism in interacting few-electron dots with 
strong spin-orbit coupling }
\author{Amin Naseri, Alex Zazunov, and Reinhold Egger}
\affiliation{Institut f\"ur Theoretische Physik,
Heinrich-Heine-Universit\"at, D-40225  D\"usseldorf, Germany}
\date{\today}
\begin{abstract} 
We study the ground state of $N$ weakly interacting electrons (with
$N\le 10$)
in a two-dimensional parabolic quantum dot with strong Rashba 
spin-orbit coupling.  Using dimensionless parameters for the
Coulomb interaction, $\lambda\alt 1$, and the Rashba coupling, $\alpha\gg 1$, 
the low-energy physics is characterized by an almost flat single-particle
dispersion.  From an analytical approach for  
$\alpha\to \infty$ and $N=2$, and from numerical
exact diagonalization and Hartree-Fock calculations, we find 
a transition from a conventional unmagnetized ground state (for
$\lambda<\lambda_c$) to an orbital ferromagnet (for $\lambda>\lambda_c$), 
with a large magnetization and a circulating charge current. 
We show that the critical interaction strength, 
$\lambda_c=\lambda_c(\alpha,N)$, vanishes in the limit $\alpha\to \infty$.
\end{abstract}
\maketitle

\section{Introduction}\label{sec1}

The electronic properties of few-electron quantum dots in 
semiconductor nanostructures have been widely studied over the past 
decades \cite{nazarov,reimann,kouwenhoven}.  
Typically, electrons in the two-dimensional (2D) electron gas  
formed at the interface between different semiconductor layers 
are confined to a localized region in space by means of electrostatic 
trapping. The resulting confinement is usually well approximated by 
a parabolic potential with oscillator frequency $\omega$,
suggesting a simple 2D oscillator spectrum. 
However, Coulomb interactions are important in such devices,
and their impact can readily be seen in transport
 spectroscopy \cite{kouwenhoven}.  Apart from the ubiquitous Coulomb 
charging effects, they are also predicted to induce a transition to a 
finite-size Wigner crystal of $N$ electrons, the ``Wigner 
molecule'' \cite{egger,landman},  where the electrostatic repulsion 
suppresses quantum fluctuations and inter-electron distances are 
maximized \cite{jauregui,peeters,peeters2}. 
The ratio between the confinement scale, $l_T=\sqrt{\hbar/m_e\omega}$,
with the effective mass $m_e$, and 
the Bohr radius, $a_B=\hbar^2\varepsilon_0/m_{e} e^2$, 
defines a dimensionless interaction strength parameter \cite{reimann},
\begin{equation}\label{lambdadef}
\lambda= \frac{l_T}{a_B}=\frac{e^2}{\varepsilon_0 \hbar\omega l_T}.
\end{equation}
Interactions are here described by the standard Coulomb 
potential, $V({\bf r})=e^2/\varepsilon_0 r$, 
where the dielectric constant $\varepsilon_0$ accounts for
static external screening.  The crossover from the weakly
interacting Fermi liquid phase (realized for $\lambda\ll 1$) 
to the Wigner molecule then happens around $\lambda\approx 1$ and
is known to be rather sharp despite of the finite-size geometry \cite{egger}.   
Due to the confinement-induced reduction of quantum fluctuations,
the corresponding electron densities near the transition are 
much higher than the one required for bulk Wigner crystal 
formation \cite{reimann}.

Another modification of the 2D oscillator spectrum is caused by 
spin-orbit coupling.  We here focus on the Rashba term 
caused by interface electric  fields, which often  
is the dominant spin-orbit coupling and can be tuned 
by gate voltages \cite{winkler}.  
Other types of spin-orbit coupling are expected to generate
similar physics as described below, assuming that one can  
reach the corresponding strong-coupling regime.  
In particular, the model studied below applies directly to the 
case of Dresselhaus spin-orbit coupling \cite{winkler}.
With the Rashba wavenumber 
$k_0$, it is convenient to employ a dimensionless Rashba coupling, 
\begin{equation}\label{alphadef}
\alpha = k_0 l_T = k_0 \sqrt{ \frac{\hbar}{m_e \omega} }. 
\end{equation}
The single-particle spectrum of a dot with weak Rashba coupling,
$\alpha\alt 1$, has been discussed, e.g., in Refs.~\cite{tretyak,tapash}.
Interaction effects in few-electron dots with $\alpha\alt 1$ 
have been investigated by density functional theory \cite{governale},
 quantum Monte Carlo simulations \cite{pede,weissegger,pederiva},
exact diagonalization \cite{ulloa0,ulloa,chakra}, and 
configuration interaction calculations \cite{cavalli}.
With increasing $\alpha$, the Wigner molecule transition was
found to shift to weaker interactions, i.e., to smaller $\lambda$.
As noted in Refs.~\cite{kivelson,silvestrov},
the related bulk Wigner crystal formation is also easier
to achieve when the Rashba term is present.

In this paper, we study interacting few-electron quantum dots in the 
regime of large Rashba spin-orbit coupling, $\alpha\gg 1$.  
This regime appears to be within close experimental reach
\cite{nitta,ensslin,ast0,ast1,ast2,aruga,tarucha,murakawa},
and is also of considerable fundamental interest.  In fact,
many materials with strong spin-orbit coupling are known to realize a 
topological insulator phase \cite{hasan,qi}.
Near the boundary of a noninteracting 2D topological 
insulator with time reversal symmetry (TRS), an odd number of 
gapless one-dimensional (1D) helical edge states must be 
present \cite{hasan,qi}, where the spin is tied to the momentum of 
the electron.  As we do not address magnetic field effects here, 
the Hamiltonian below enjoys TRS. Moreover, it is characterized 
by strong spin-orbit coupling, and it resembles a topological 
insulator in the absence of interactions.

Given the above developments, it is not surprising that 
several theoretical works \cite{li1,li2,rashba,ghosh,sherman} have 
already addressed the physics of noninteracting electrons in quantum dots with 
$\alpha\gg 1$.   In this limit, the low-energy spectrum of a 
parabolic dot is well described by a sequence of almost flat 
Landau-like bands (see Sec.~\ref{sec2a}),
\begin{equation}\label{dispersionrelation}
E_{J,n} \simeq \hbar\omega \left( n 
+ \frac12  + \frac{J^2}{2 \alpha^2} \right) ,
\end{equation}
with half-integer total angular momentum $J$ and the 
band index $n=0,1,2,\ldots$, such that states with the same 
$n$ but different $J$ are almost degenerate.
Equation (\ref{dispersionrelation}) reflects the spectrum of a 1D 
(radial) oscillator plus a decoupled rotor with large moment of inertia.
Assuming that the Fermi energy is within the $n=0$ band,
with corresponding Fermi angular momentum $J_F$, the Kramers pair with 
$J=\pm J_F$ has eigenfunctions localized near the ``edge'' of the dot. 
In fact, those states have the largest distance from the dot center among 
all occupied states, and form a helical edge
with opposite spin orientation of the counterpropagating $\pm J_F$ states 
\cite{li1}.  By virtue of the bulk-boundary correspondence \cite{hasan}, 
the authors of Ref.~\cite{li1} argued that a noninteracting dot with 
$\alpha\gg 1$ has features similar to the finite-size version
of a 2D topological insulator. Indeed,  time reversal invariant 
single-particle perturbations, e.g., representing the effects of 
elastic disorder, are predicted not to mix opposite-spin states, 
and the helical edge is therefore protected against such sources of 
backscattering.   In the finite-size dot geometry, however, 
the $\mathbb{Z}_2$ invariant commonly employed to classify 
the topological insulator phase is not well defined.

For a dot with $\alpha\gg 1$, since the noninteracting spectrum is 
almost flat, one can expect that interactions have a profound
effect.  For instance, in lattice models 
hosting a topological insulator phase for weak interactions,
Mott insulator or spin liquid phases emerge for strong
interactions \cite{balents}; for the case of interacting bosons, 
see Refs.~\cite{atoms1,atoms2,atoms3,atoms4,atoms5}. 
Moreover, the conspiracy of a single-particle potential with sufficiently 
strong Coulomb interactions can induce two-particle
Umklapp processes destroying the helical edge state \cite{xumoore,wuzhang}.  
Motivated by these developments, we here study the ground state of 
interacting electrons in a quantum dot with strong Rashba spin-orbit coupling.  
We find it quite remarkable that the relatively simple Hamiltonian
below captures such diverse behaviors as Wigner molecule formation, 
the presence of helical edge states, and -- as we shall
argue -- the molecular equivalent of an \textit{orbital ferromagnet}.
This Hamiltonian is also expected to accurately describe semiconductor 
experiments, where recent progress holds promise of reaching the ultra-strong
Rashba coupling regime.  Let us now briefly summarize our main 
results, along with a description of the structure of the paper.

In Sec.~\ref{sec2a}, we present the single-particle model for the quantum dot,
and summarize its solution for large Rashba parameter $\alpha$.
While our general conclusions hold for arbitrary radially symmetric
confinement, quantitative results are provided for the most
important case of a parabolic trap.  
We introduce a single-band approximation valid for weak-to-intermediate
interaction strength, $\lambda\alt 1$, and energy scales below $\hbar\omega$,
which  allows one to make significant analytical progress.
In Sec.~\ref{sec2b}, we then discuss the general properties of 
Coulomb matrix elements.  The limit of ultra-strong Rashba 
coupling, $\alpha\to \infty$,  is addressed in Sec.~\ref{sec2c}, 
where a simple analytical result for the Coulomb matrix elements 
is derived.  For the resulting $\alpha\to \infty$ model, $H_\infty$,
already weak interactions induce strongly correlated phases.
The Coulomb matrix elements not included in $H_{\infty}$,
arising for large but finite $\alpha$, 
are addressed in detail in Sec.~\ref{sec2d}.

Next, in Sec.~\ref{sec3}, we present the exact ground-state solution of
$H_\infty$ for two electrons ($N=2$). 
While the above discussion may suggest that a Wigner molecule 
will be formed, we find an orbital ferromagnetic state.
The $N=2$ ground state of $H_\infty$, see Sec.~\ref{sec3a}, is shown to be
highly degenerate in Sec.~\ref{sec3b}.  However, perturbative inclusion of 
Coulomb corrections beyond $H_{\infty}$, see Sec.~\ref{sec3c}, 
breaks the degeneracy and suggest the possibility of spontaneously broken
TRS in an interacting $N=2$ dot (for a more precise characterization 
of this phenomenon, see Sec.~\ref{sec3}), with a large value of the 
total angular momentum found already for weak interactions.
The emergence of a finite magnetization \cite{foot01}, $M_s\ne 0$, 
suggests a finite-size (``molecular'') version of an orbital ferromagnet. 
This remarkable behavior appears at arbitrarily weak (but finite) 
interaction strength, with giant values of the  magnetization. 
We estimate $M_s\approx (\lambda\alpha)^{1/4}\hbar$, see Sec.~\ref{sec3c}. 
This highlights that the orbital angular momentum is behind this phenomenon,
see also Ref.~\cite{hernando}.  

In Sec.~\ref{sec4a}, we then present exact diagonalization results for the 
ground-state energy of $N=2$ and $N=3$ electrons in the dot for 
$\alpha=10$ and $\alpha=15$, going beyond the $\alpha\to\infty$ model 
$H_\infty$.  We now find that only above a critical interaction strength, 
$\lambda>\lambda_c(\alpha,N)$, the dot develops a magnetization, 
 $M_s\ne 0$.  The parameter $\lambda_c$ becomes smaller
with increasing $\alpha$, which is consistent with 
 $\lambda_c(\alpha\to \infty)\to 0$ as obtained from $H_\infty$
 in Sec.~\ref{sec3}.  In Sec.~\ref{sec4b}, we then 
discuss Hartree-Fock (HF) results
for particle numbers up to $N=10$, where exact diagonalization 
becomes computationally too expensive.  The HF results show qualitatively
the same effects, indicating that orbital ferromagnetism represents the
generic behavior of weakly interacting electrons in quantum dots with 
ultra-strong Rashba coupling.  Finally, we conclude in Sec.~\ref{sec5},
where we also discuss perspectives for experiments. 
Additional details about the $\alpha\to\infty$ limit are given in an Appendix.

\section{Coulomb interactions in a Rashba dot}\label{sec2}

\subsection{Single particle problem}\label{sec2a}

We consider electrons in a 2D quantum dot with 
parabolic confinement in the $xy$ plane. Including the Rashba 
spin-orbit coupling, the single-particle Hamiltonian reads \cite{winkler} 
\begin{equation} \label{H}
H_{\rm dot} = \frac{\hbar^2 }{2m_e} 
{\bf k}^2 + \frac{m_e}{2} \omega^2 {\bf r}^2 -
\frac{\hbar^2 k_0 k}{m_e} {\cal P}_h ,
\end{equation}
where ${\bf k} = -i(\partial_x,\partial_y)$, 
${\bf r} = ( x,y)$, $\omega$ is the trap frequency (defined in the
absence of spin-orbit coupling), and
the positive wavenumber $k_0$ determines the Rashba coupling.
With Pauli matrices $\sigma_{x,y,z}$ referring to the electronic spin, 
the Hermitian helicity operator, ${\cal P}_h = 
(k_y\sigma_x-k_x\sigma_y)/k$, has the eigenvalues $\pm 1$.  
In the absence of the trap ($\omega=0$), helicity and 
momentum are conserved quantities.   Writing  
${\bf k} =  k ( \cos \phi, \sin \phi)$, it is a simple 
exercise to obtain the ${\cal P}_h$-eigenspinors,
$\Phi_\pm(\phi)$, with conserved helicity $\pm 1$.
The dispersion relation is then (up to a constant shift) 
given by $\hbar^2
(k \mp k_0)^2/2m_e$.  Low-energy states have positive helicity with
\begin{equation}
\Phi_+ (\phi) = \frac{1}{\sqrt{2}} \left(\begin{array}{c} 1\\
-i e^{i \phi}\end{array}\right),
\end{equation}
and for given $k\approx k_0$, a U(1) degeneracy is realized, 
corresponding to a ring in momentum space. 

In the presence of the trap, however, helicity and momentum are not 
conserved anymore. The system now has two characteristic length scales, 
namely the confinement scale, $l_T=\sqrt{\hbar/m_e\omega}$, and the 
spin-orbit length, $1/k_0$.  
Their ratio determines the dimensionless Rashba parameter $\alpha$ 
in Eq.~\eqref{alphadef}.
In this paper, we discuss the case $\alpha\gg 1$, where  
positive helicity states are separated from ${\cal P}_h=-1$ states 
by a huge gap of order 
$\hbar^2 k_0^2/m_e =\alpha^2\hbar\omega$.
As a consequence, negative helicity states can safely be projected away.
Noting that the total angular momentum operator, $J_z=-i\hbar\partial_\phi+
\hbar\sigma_z/2$, is conserved, with eigenvalues 
$\hbar J$ (half-integer $J$), the low-energy eigenstates of 
$H_{\rm dot}$ for $\alpha\gg 1$ have the momentum representation 
\begin{equation}\label{wavef}
\psi_{J,n}(\kappa,\phi) = \frac{e^{i(J-1/2)\phi}}{\sqrt{2\pi \kappa}} 
\ u_{J,n}(\kappa) \Phi_+(\phi),
\end{equation}
where we use the dimensionless positive wavenumber $\kappa=k l_T$. 
The radial wavefunction, $u_{J,n}(\kappa)$, 
obeys the effective 1D Schr\"odinger equation \cite{li1,ghosh} 
\begin{equation}\label{radialH}
\left( - \frac12\partial_\kappa^2 +
\frac12 (\kappa - \alpha)^2 + \frac{J^2}{2 \kappa^2} - \frac{E_{J,n}}{
\hbar\omega} \right) u_{J,n}(\kappa) =  0,
\end{equation}
where $n=0,1,2,\ldots$ labels the solutions.
For $\alpha \gg 1$, it is justified to approximate Eq.~\eqref{radialH} by 
replacing $J^2/2\kappa^2 \to J^2/2\alpha^2$. 
The radial problem then decouples from the angular one and 
becomes equivalent to a shifted 1D oscillator with 
energy levels $(n+1/2) \hbar \omega$.  
Moreover, the angular problem reduces to a rigid rotor with the
large moment of inertia $\alpha^2/\hbar\omega$.  
We thus arrive at the $E_{J,n}$ quoted in Eq.~(\ref{dispersionrelation}),
where $n$ serves as band index and $J$ labels  
the almost degenerate states within each band.  
We find that corrections to the energies in 
Eq.~\eqref{dispersionrelation} scale $\sim 1/\alpha^{3}$ for $\alpha\gg 1$. 
In fact, a recent numerical study of $H_{\rm dot}$ 
has reported that Eq.~(\ref{dispersionrelation}) is
highly accurate for $\alpha\agt 4$ \cite{ghosh}.

For weak-to-intermediate Coulomb interaction strength, only low-energy
states are needed to span the effective Hilbert space 
determining the ground state.  
It can then be justified to retain only $n=0$ modes.
This step implies the restriction to angular momentum states
with $|J|  \alt \alpha$, since otherwise $n>0$ states should also be included. 
For the results below, we have checked
that this ``single-band approximation'' is indeed justified.
From now on, the single-particle Hilbert space is restricted 
to the $n=0$ sector (and the $n$ index will be dropped). 
In momentum representation, this space is spanned by
the orthonormal set of states  \cite{footnew}
\begin{equation}\label{psiJ}
\psi_{J}(\kappa,\phi) = \frac{\pi^{1/4}l_T}{\sqrt{\kappa}} 
e^{-(\kappa-\alpha)^2/2} e^{i(J-1/2)\phi} \left( 
\begin{array}{c} 1 \\ -i e^{i \phi} \end{array} \right).
\end{equation}
Up to the zero-point contribution, the corresponding single-particle energy
is $E_J=  J^2\hbar\omega/2\alpha^2$.  The momentum-space 
probability density for all states is independent of $J$,  
representing a radially symmetric Gaussian peak centered at $k=k_0$,
\begin{equation} \label{radsym}
\rho_J(k)= |\psi_J|^2 = \frac{2\sqrt{\pi} l_T}{k} e^{-(k-k_0)^2l_T^2}.
\end{equation}
The coordinate representation of Eq.~(\ref{psiJ}) now follows by
Fourier transformation,
\begin{eqnarray}\nonumber
&& \tilde\psi_{J}(\rho,\theta) = \frac{i^{J-1/2}}{l_T} e^{i (J-1/2) \theta}
\left( \begin{array}{c} F_{J-1/2}(\rho) \\ 
e^{i \theta} F_{J+1/2}(\rho) \end{array} \right), \\
\label{Fm} && F_{m}(\rho) =\int_0^\infty 
\frac{ d\kappa \sqrt{\kappa}}{2\pi^{3/4}} 
e^{-(\kappa-\alpha)^2/2} J_m(\kappa \rho),
\end{eqnarray}
where ${\bf r} = r (\cos \theta, \sin \theta)$ with
$\rho=r/l_T$, and we use the Bessel functions $J_m(x)$ (integer $m$).

It will be convenient to use a second-quantized formalism below, with the  
noninteracting Hamiltonian 
\begin{equation}\label{h02q}
H_0 = \sum_J E^{}_J c_J^{\dagger} c_J^{}, \quad E_J=\frac{J^2}{2\alpha^2}\hbar
\omega,
\end{equation}
where fermion annihilation operators are denoted by $c_J^{}$ for 
half-integer $J$. The electron field operator is then given by
\begin{equation}\label{field}
\Psi({\bf r}) =  \sum_{J} \tilde\psi_J({\bf r}) c_J .
\end{equation}
The noninteracting ground state is a Fermi sea with all 
states $|J|\le J_F \alt \alpha$ occupied. 
For even number of electrons in the dot, $N=2J_F+1$,
the ground state is unique and has the energy
$E_0=N (N^2-1)\hbar \omega/24\alpha^2$.
When $N$ is odd, however, the ground state is two-fold degenerate.  
Note that the single-band approximation can  only be 
justified for $N\alt \alpha$. 

Next, we introduce the total angular momentum operator
of the interacting $N$-electron dot,
\begin{equation} \label{opms}
\hat M_s= \hbar \sum_J J c_J^\dagger c_J^{},
\end{equation}
which is conserved even in the interacting case. 
Noting that the Hamiltonian respects TRS, a finite ground-state 
expectation value, $M_s=\langle \hat M_s\rangle \ne 0$, 
corresponds to a spontaneous magnetization of the dot and thus would
imply that the ground state breaks TRS.  
For the noninteracting case, recent work has discussed a spin-orbit-induced 
orbital magnetization in similar nanostructures, either in the 
presence \cite{mishch} or absence \cite{flatte} of a magnetic Zeeman field.
We find below that, in the absence of a magnetic field but with 
strong spin-orbit coupling, already weak
interactions can induce a transition to an orbital ferromagnet,
where a large magnetization is present and
the electrons in the dot carry a circulating charge current.
This behavior appears for $\lambda>\lambda_c$,
where the critical interaction strength, $\lambda_c$,
vanishes in the limit $\alpha\to \infty$.

\subsection{Coulomb matrix elements} \label{sec2b}

The second-quantized Hamiltonian, $H=H_0+H_I$, with $H_0$ 
in Eq.~\eqref{h02q}, includes a normal-ordered Coulomb interaction term, 
\begin{equation} \label{Hint2}
H_{I} = \frac12 \int d^2 {\bf r}_1 d^2 {\bf r}_2 \, 
V({\bf r}_1 - {\bf r}_2) \Psi^\dagger({\bf r}_1) 
\Psi^\dagger({\bf r}_2) \Psi({\bf r}_2) \Psi({\bf r}_1),
\end{equation}
where $V({\bf r})$ is the Coulomb potential.
Inserting the field operator \eqref{field}, 
and taking into account angular momentum conservation, we find  
\begin{equation}\label{hintfin}
 H_{I} =\sum_{J_1, J_2;m } V_{J_1, J_2}^{(m)} \,
c^\dagger_{J_1+m} c^\dagger_{J_2-m} c^{}_{J_2} c^{}_{J_1} ,
\end{equation}
with the integer angular momentum exchange $m$.
The real-valued Coulomb matrix elements in Eq.~\eqref{hintfin} 
take the form
\begin{eqnarray}\label{Vm1m2m}
&& V_{J_1 ,J_2}^{(m)} = 2\pi\lambda \hbar\omega
\int_0^\pi d\theta \, \cos (m \theta) \\ && \times~~
\nonumber \int_0^\infty d\rho \int_0^\infty d\rho' \frac{G_{J_1,J_1+m}(\rho)
 G_{J_2,J_2-m}(\rho') } 
{\sqrt{\rho^2 + \rho^{\prime 2} - 2 \rho \rho' \cos \theta}} ,
\end{eqnarray}
where we define 
\begin{equation}\label{Gdef}
G_{J,J'}(\rho) = \rho \sum_{\sigma=\pm} 
F_{J+\sigma/2}(\rho) F_{J'+\sigma/2}(\rho) = G_{J',J}(\rho)
\end{equation}
with $F_m(\rho)$ in Eq.~\eqref{Fm}.  Using a well-known 
expansion formula,
\begin{equation}
\frac{1}{\sqrt{\rho^2+\rho^{\prime 2}-2\rho\rho'\cos\theta}}
= \frac{1}{\rho_>}
\sum_{l=0}^\infty \left(\frac{\rho_<}{\rho_>}\right)^l P_l(\cos\theta) ,
\end{equation}
where $\rho_>$ ($\rho_<$) is the larger (smaller) of $\rho$ and $\rho^\prime$,
the denominator in Eq.~\eqref{Vm1m2m} is expressed as a series involving
Legendre polynomials, $P_l(x)$.  This allows us to perform the 
$\theta$-integral in Eq.~\eqref{Vm1m2m} 
analytically, and after some algebra we obtain
\begin{eqnarray}\nonumber
V_{J_1, J_2}^{(m)} &=&  2\pi^2 \lambda \hbar \omega
\sum_{l=|m|,|m|+2,\cdots} {\cal R}_l^{(m)} 
\int_0^\infty \frac{d\rho}{\rho^{l+1}} \int_0^{\rho} \rho^{\prime l} 
 d\rho'\\ &\times&
\label{Vmm} \left [ G_{J_1,J_1+m}(\rho) G_{J_2,J_2-m}(\rho')+
(\rho\leftrightarrow\rho') \right], 
\end{eqnarray} 
with the numbers (see also Ref.~\cite{jpa})
\begin{equation}\label{r2def}
{\cal R}_l^{(m)} = \frac{(2l-1)!!}{2^l l!} \prod_{n=1}^{(l+|m|)/2} 
\frac{(n-1/2) (l-n+1)}{ n (l-n+1/2)}
\end{equation}
and ${\cal R}_0^{(0)}=1$.  The Coulomb matrix elements in Eq.~\eqref{Vmm}
are in a convenient form for numerics \cite{foot1}.  In addition,
as we discuss next, Eq.~\eqref{Vmm} also allows for analytical progress 
in the limit $\alpha\to\infty$.

\subsection{Ultra-strong Rashba coupling}
\label{sec2c}

The interaction matrix elements (\ref{Vmm}) can be computed in closed 
form for $\alpha\to \infty$.  For consistency with the
 single-band approximation, this limit is taken as $k_0\to \infty$
with $l_T$ held finite, i.e., we  assume ultra-strong Rashba coupling 
in the presence of the dot.  Taking the limit in opposite order 
gives similar but slightly different results; we provide a 
discussion of this point in the Appendix.  

For $\alpha\to \infty$, using Eq.~(\ref{Fm}) and $\rho=r/l_T$,
the single-particle states have the asymptotic real-space representation 
\begin{equation}\label{wlar}
\tilde\psi_{J}(\rho,\theta) \simeq \frac{i^{J-1/2}
e^{i(J-1/2)\theta} e^{-\rho^2/2} }{\pi^{3/4}l_T\sqrt{\rho}} 
\left(\begin{array}{c} \cos \left(\alpha\rho-\frac{\pi J}{2}\right) 
\\ e^{i\theta} \sin\left(\alpha\rho-\frac{\pi J}{2}\right)\end{array} \right),
\end{equation}
where the Gaussian $e^{-\rho^2/2}$ factor reflects the trap potential and 
the Rashba coupling causes rapid oscillations.   
Equation \eqref{Gdef} is well-defined in the $\alpha\to \infty$ limit, 
$G_{J,J+m}(\rho) \to \pi^{-3/2}e^{-\rho^2} \cos(\pi m/2)$.
Notably, for odd $m$, we find $G=0$, leading
to the even-odd parity effect described below.
Performing the remaining integrations in Eq.~(\ref{Vmm}), we obtain a
surprisingly simple result for the Coulomb matrix elements,
\begin{equation}\label{vmla}
\lim_{\alpha\to \infty} V_{J_1,J_2}^{(m)} = \lambda\hbar \omega  S_m .
\end{equation}
In terms of the ${\cal R}_l^{(m)}$ in Eq.~(\ref{r2def}), 
the numbers $S_m=S_{-m}$ are nonzero only for even $m$,
\begin{equation}\label{Smdef}
S_m= \delta_{m,{\rm even}} \sum_{l = |m|, |m|+2, \cdots}  e^{-\eta l} 
{\cal R}_l^{(m)} C_l,
\end{equation}
with the coefficients
\begin{equation}\label{cdef}
C_l =\frac{2}{\sqrt{\pi}} \int_0^{\pi/4} d\phi \frac{\tan^l \phi}{\cos\phi}.
\end{equation}
The small parameter $\eta\ll 1$ in Eq.~\eqref{Smdef} (we take $\eta=0.01$
for concreteness below)
regularizes the $l$-summation, which for $\eta=0$ is 
logarithmically divergent with respect to the upper limit.
In physical terms, this weak divergence
comes from the singular $r\to 0$ behavior of the $1/r$ Coulomb potential, 
which in practice is cut off by the 
transverse (2D electron gas) confinement. Expressing the corresponding 
length scale as $\eta l_T$, we arrive at the regularized 
form in Eq.~\eqref{Smdef}.  Numerical results for the $S_m$  
are shown in Table~\ref{tab1}: $S_m$ has a maximum for $m=0$ and 
then decays with increasing $|m|$ \cite{footsm}.  

\begin{table}
\caption{\label{tab1} Nonvanishing $S_m$ for  $|m|\le 16$ 
from Eqs.~(\ref{Smdef}) and  (\ref{cdef}).
}
\begin{tabular}{|l|l|}
\hline
$m$ & $S_m$  \\ \hline 
0 & 1.11757 \\
2 &  0.172844 \\
4 &  0.0862971 \\
6 &  0.0556035 \\
8 &  0.0401376 \\ 
10 & 0.0309001 \\
12 & 0.0247964 \\
14 & 0.0204838 \\
16 & 0.0172877 \\ \hline
\end{tabular}
\end{table}

It is worth pointing out that the $\alpha\to \infty$ Coulomb matrix elements 
in Eq.~\eqref{vmla} are valid for \textit{arbitrary} 
radially symmetric confinement, where different confinement potentials only lead
to different coefficients $C_l$. While Eq.~\eqref{cdef} describes the 
parabolic trap, taking for instance a hard-wall circular confinement
\cite{gogolin}, we find $C_l = 4/[\pi (l+1)]$.

An important consequence of Eq.~\eqref{Smdef} is 
that all Coulomb matrix elements with odd $m$ vanish identically.
Equation \eqref{vmla} therefore predicts a pronounced
 \textit{parity effect}: 
Depending on the parity of the exchanged angular momentum $m$, 
$V^{(m)}_{J_1,J_2}$ is either finite or zero.
Another important feature is that the $V^{(m)}_{J_1,J_2}$
in Eq.~\eqref{vmla} are completely independent
of the ``incoming'' angular momenta $J_1$ and $J_2$.
This can be rationalized by noting that in the $\alpha\to \infty$ limit, we 
arrive at an effectively homogeneous 1D problem
corresponding to a ring in momentum space, see also the Appendix.  
For a homogeneous electron gas, on the other hand, it is well known that
interaction matrix elements only depend on the exchanged 
(angular) momentum but not on particle momenta themselves \cite{giul}.
With $H_0$ in Eq.~\eqref{h02q}, the conserved particle number,
$N=\sum_J c_J^\dagger c_J^{}$, and noting that
$S_m=0$ for odd $m$, the $\alpha\to \infty$ Hamiltonian takes the form
\begin{equation}\label{universalH}
H_\infty =\lambda\hbar\omega \sum_{m\ne 0}  S_m \sum_{J_1,J_2}  
c^\dagger_{J_1+m} c^\dagger_{J_2-m} c^{}_{J_2} c^{}_{J_1} +H_0+E_s,
\end{equation}
with the energy shift $E_s=S_0 N(N-1)\lambda\hbar\omega$. 
Since $S_0$ enters only via this energy shift, but otherwise disappears
in $H_\infty$, it is convenient to put $S_0=0$ from now on and let
the sum in Eq.~\eqref{universalH} include $m=0$;
the energy $E_s$ will be kept implicit in what follows.  
Corrections to $H_\infty$ at finite $\alpha$ originate from 
Coulomb matrix element contributions that vanish for $\alpha\to \infty$,
in particular those with odd $m$. 
In Sec.~\ref{sec3}, we shall discuss the exact ground state of  
 $H_\infty$ for $N=2$.

\subsection{General properties of Coulomb matrix elements}
\label{sec2d}

We proceed by presenting symmetry relations  
relating different Coulomb matrix elements in Eq.~(\ref{Vmm}). 
Note that our discussion here is not restricted to 
$\alpha\to \infty$, but applies to finite Rashba couplings with $\alpha
\gg 1$.  First, by virtue of particle indistinguishability, 
\begin{equation} \label{srel1}
V_{J_1, J_2}^{(m)} = V_{J_2 ,J_1}^{(-m)} .
\end{equation}
Additional symmetry relations follow from the time reversal invariance
of the interaction Hamiltonian $H_I$. Indeed, because of TRS,
Eq.~\eqref{Gdef} yields $G_{-(J+1),-(J+1+m)}(\rho) = 
(-1)^m G_{J,J+m}(\rho)$, which then leads to the symmetry relations
\cite{footsym}
\begin{eqnarray} \label{Vsymmrel3}
V_{J_1, J_2}^{(m)} & = & V_{-J_1, -J_2}^{(-m)} \\ \nonumber
 &=&  (-1)^m V_{-J_1-m, J_2}^{(m)} = 
(-1)^m V_{J_1, -J_2+m}^{(m)} .
\end{eqnarray}
In particular, for odd $m$ and arbitrary $J$, Eq.~\eqref{Vsymmrel3} yields
\begin{equation} \label{weakparity}
V_{-m/2, J}^{(m)} = V_{J, m/2}^{(m)} = 0 .
\end{equation}
The parity effect found in Sec.~\ref{sec2b} for $\alpha\to \infty$, 
with $V^{(m)}_{J_1,J_2}=0$ for all odd $m$, 
is consistent with Eq.~(\ref{weakparity}):
While the finite-$\alpha$ relation (\ref{weakparity}) only implies that 
certain odd-$m$ matrix elements have to vanish,
the $(J_1,J_2)$-independence of the Coulomb matrix elements 
in the limit $\alpha\to\infty$ forces all of them to vanish for odd $m$.   
Finally, numerical calculation of the $V^{(m)}_{J_1,J_2}$
can take advantage of Eqs.~(\ref{srel1}) and (\ref{Vsymmrel3}), 
since all Coulomb matrix elements follow
from the knowledge of $V_{J_1 J_2}^{(m\ge 0)}$ with
$J_{1,2} \geq 1 \mp m/2$ when $m$ is odd, and 
$J_{1,2} \geq (1 \mp m)/2$ when $m$ is even.

\begin{figure}
\begin{center}
\includegraphics[width=0.5\textwidth]{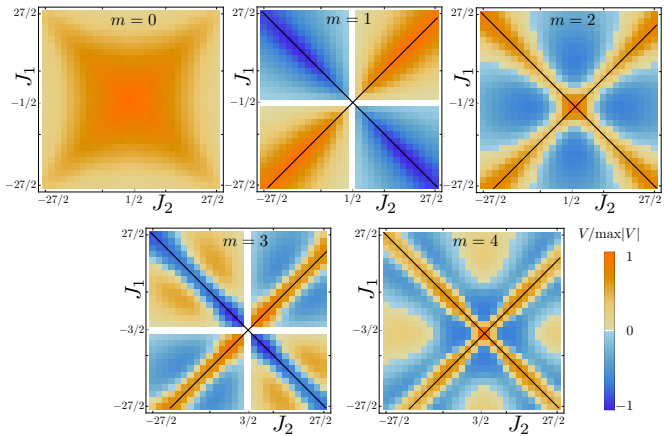}
\end{center}
\caption{\label{fig1}  
Color-scale plot of the Coulomb matrix elements 
$V_{J_1 ,J_2}^{(m)}$ in the $J_1$-$J_2$ plane, normalized to their
maximum value in the shown region, 
for Rashba parameter $\alpha = 10$ and various $m$. 
Matrix elements taken along the solid lines are shown in Fig.~\ref{fig2}.}
\end{figure}

\begin{figure}
\begin{center}
\includegraphics[width=0.55\textwidth]{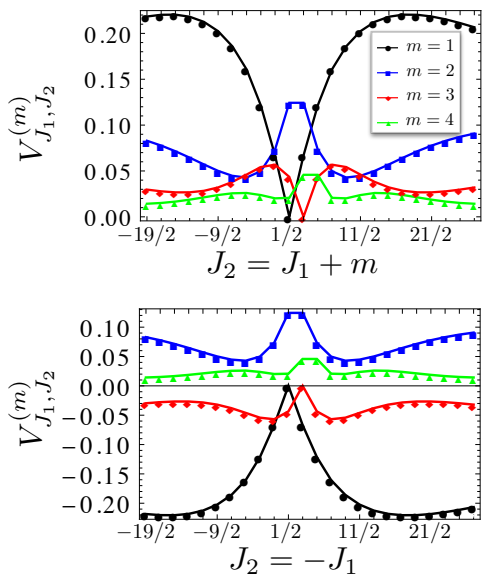}
\end{center}
\caption{\label{fig2} Coulomb matrix elements,  
in units of $\lambda\hbar\omega$, vs angular momentum $J_2$.  The plots are 
for $\alpha=10$ and various $m$.
The upper (lower) panel shows the case $J_2=J_1+m$ ($J_2=-J_1$), 
resp.,  cf.~the solid lines in Fig.~\ref{fig1}. }
\end{figure}

Numerical results for $\alpha=10$ and several $m$ are shown in 
Figs.~\ref{fig1} and \ref{fig2}.  We draw the following conclusions:
\begin{itemize}
\item With increasing $|m|$, the absolute magnitude of the 
Coulomb matrix elements quickly decreases.  
\item Pronounced differences between even and odd $m$ 
are not yet visible for $\alpha=10$.  Additional calculations 
for $\alpha=15$ and $\alpha=30$ (not shown here) confirm
that the matrix elements for odd $m$ become
more and more suppressed relative to the even-$m$ case.  
However, the ideal parity effect, where all odd-$m$ matrix 
elements vanish for $\alpha\to \infty$, is approached rather slowly. 
\item For $\alpha=10$, Figs.~\ref{fig1} and \ref{fig2} show that the 
$V_{J_1, J_2}^{(m)}$ carry a significant
dependence on the indices $(J_1,J_2)$. 
This dependence ultimately disappears for $\alpha\to \infty$.
\item
For given value of $m$, the matrix element $V^{(m)}_{J_{1}J_{2}}$ 
has maximal absolute magnitude along the two lines $J_{2}=-J_{1}$ and
$J_{2}=J_{1}+m$ in the $(J_{1},J_{2})$ plane.
Noting that the single-particle eigenfunctions are localized near a 
ring of radius $k_0$ in momentum space, these two lines can be 
interpreted as BCS-like and exchange-type scattering processes, respectively, 
cf.~the Appendix.
The two lines of maximal absolute magnitude are orthogonal to each other,
and cross at the point $(-m/2,m/2)$ in 
the ($J_1,J_2$) plane.  While for even $m$, this point is not a
physically realized one (since $J_{1,2}$ must be half-integer), it is always 
the  symmetry center.
\item 
$V^{(m)}_{J_{1}J_{2}}$ is positive definite along the line $J_{2}=J_{1}+m$,
for both even and odd $m$, while it is negative (positive) definite 
along the line $J_{2}=-J_{1}$ for odd (even) $m$. 
\item For even $m$, the interaction matrix elements are maximal 
at the four points where $(J_1,J_2)$ is either  given by
$\left(-\frac{m\pm 1 }{2}, \frac{m\pm 1}{2}\right)$
or by $\left(-\frac{m\pm 1 }{2}, \frac{m\mp 1}{2}\right)$.
For odd $m$, the matrix elements vanish along the lines $J_1=-m/2$ and
$J_2=m/2$, 
in accordance with the symmetry relation \eqref{weakparity}.  
\end{itemize}

\section{Two interacting electrons for
ultra-strong Rashba coupling} \label{sec3}

In this section, the $\alpha\to \infty$ model, $H_\infty$ 
[Eq.~\eqref{universalH}], is studied for $N=2$ electrons.
$H_\infty$ neglects all Coulomb matrix element contributions
beyond Eq.~\eqref{vmla}, but includes the kinetic term $H_0$.  
We assume that the interaction strength is finite,
 but $\lambda\alt 1$ is needed to validate
 the single-band approximation.

\subsection{Two-particle eigenstates}
\label{sec3a}

\begin{figure}
\begin{center}
\includegraphics[width=0.55\textwidth]{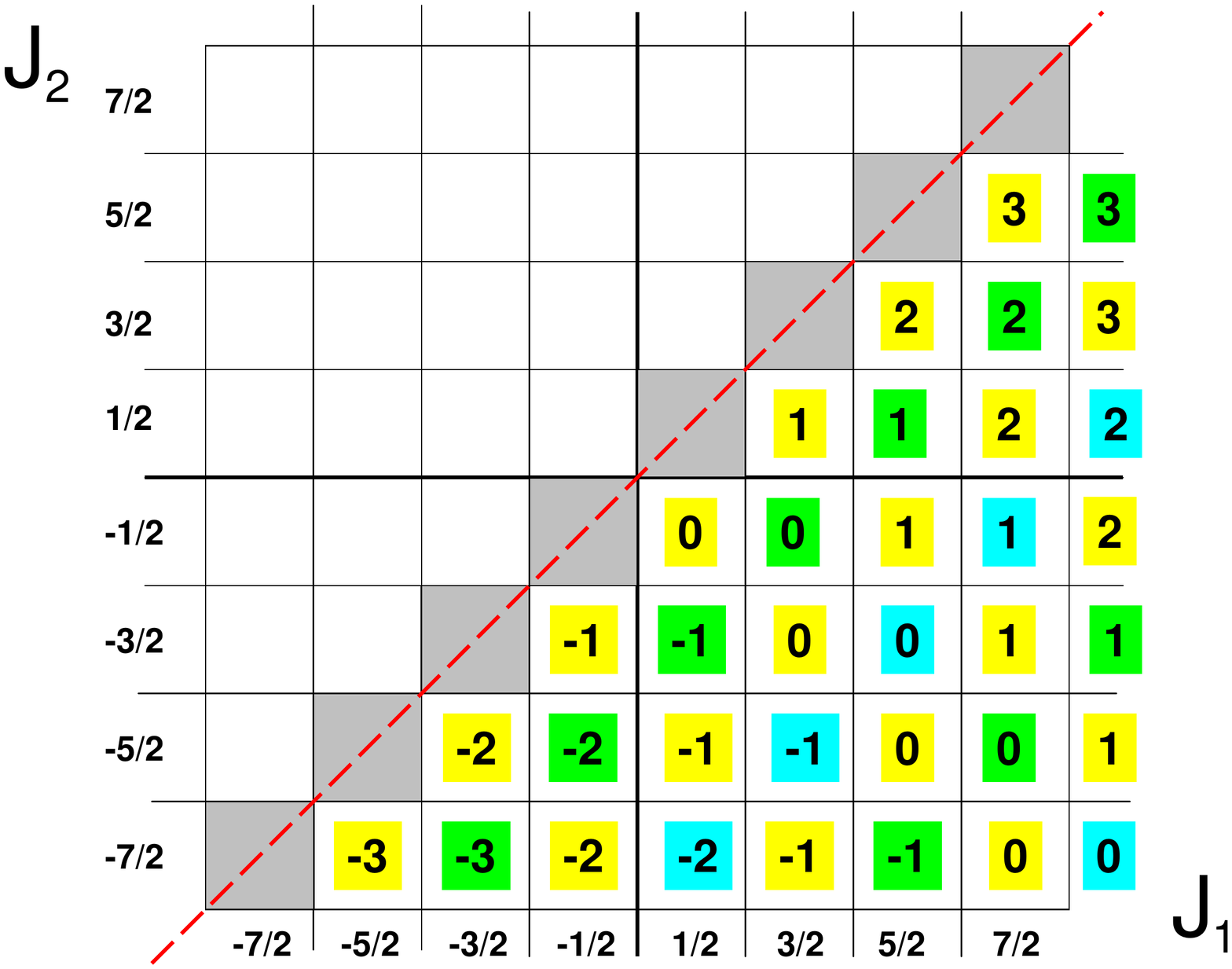}
\end{center}
\caption{\label{fig3} 
Schematic illustration of the invariant two-particle states $|M,\gamma\rangle$
(with integer $M$ and ``family'' index $\gamma=1,2,3$), see 
Eq.~\eqref{2particle}, in the $J_1$-$J_2$ plane. These states
span the complete two-particle Hilbert space.
Our ordering convention, $J_1>J_2$, implies that
only states below the main diagonal (dashed red line) appear.
Yellow cells correspond to $\gamma=1$, where the respective 
number indicates $M$.  Green (blue) cells refer to $\gamma=2$ ($\gamma=3$).
The interacting ground state has $\gamma=1$.   }
\end{figure}

The two-particle Hilbert space is spanned by 
$c_{J_1}^\dagger c_{J_2}^\dagger |0\rangle$,
where we set $J_1>J_2$ to avoid double counting and
$|0\rangle$ is the $N=0$ state. This space is composed
of decoupled subspaces, which are invariant under the action
of $H_\infty$.  The corresponding states, $|M,\gamma\rangle$,
 are labeled by the integer $M$ 
and a ``family'' index $\gamma=1,2,3$, see Fig.~\ref{fig3} for an illustration.
With amplitudes $\beta_{J>0}$ subject to the
normalization condition  $\sum_{J>0} \left|\beta_J\right|^2=1$,
and employing an auxiliary index $i_\gamma$ with values
 $i_{\gamma=1}=0$ and $i_{\gamma=2,3}=1$, those states are defined as
\begin{equation}\label{2particle}
|M,\gamma\rangle  =  \sum_{J>0} \ \beta_J  \
c_{J+M+i_{\gamma}}^\dagger c_{-J+M}^\dagger |0\rangle,
\end{equation}
where for $\gamma=2$ ($\gamma=3$), only even (odd) $J+1/2$  
are included in the summation.  

Using the energies $E_J$ [Eq.~\eqref{h02q}], some algebra shows that
the action of $H_\infty$ on such a state yields
\begin{eqnarray}  \nonumber
H_\infty|M,\gamma\rangle &=&  \sum_{J,J'>0}  
\Bigl[  \left (E_{J+M+i_\gamma}+E_{-J+M} \right)\delta_{JJ'}  \\
\label{tildebeta}
&+&   2\lambda \hbar \omega  \left ( 
S_{J-J'} - \delta_{\gamma,1} S_{J+J'}\right) \Bigr ]\\ &\times&
\nonumber  \beta_{J'}^{}\
 c_{J+M+i_\gamma}^\dagger c_{-J+M}^\dagger |0\rangle,
\end{eqnarray} 
with $S_0=0$ (see above).  Equation \eqref{tildebeta} confirms that each 
family of states stays invariant under $H_\infty$.
When looking for the ground-state energy, we note that
an $M$-dependence  can only originate from the $E_J\sim 1/\alpha^2$ terms. 
For $\alpha\to  \infty$, all $|M,\gamma\rangle$ states with 
different $M$ but the same $\gamma$, therefore, have the same energy.
As a consequence, the interacting ground state is highly degenerate 
for $\alpha\to \infty$.  This degeneracy is only
lifted by finite-$\alpha$ corrections resulting from 
the kinetic energy and from Coulomb matrix elements beyond $H_\infty$.

Importantly, since the energy-lowering contribution $-S_{J+J'}$ 
is absent in Eq.~\eqref{tildebeta} for $\gamma\ne 1$, 
the ground state must be in the $\gamma=1$ sector.  
The $\gamma=2,3$ states are separated
by an energy gap $\sim \lambda\hbar\omega$, and
we neglect these higher energy states from now on
 (and omit the $\gamma$ index). 
Since the magnetization operator $\hat M_s$ in Eq.~\eqref{opms} is conserved,
the $|M\rangle$ states are also magnetization eigenstates. Indeed, one 
immediately finds that the corresponding 
eigenvalue is $M_s=2M\hbar$.
 
\subsection{Distribution function}
\label{sec3b}

\begin{figure}
\begin{center}
\includegraphics[width=0.48\textwidth]{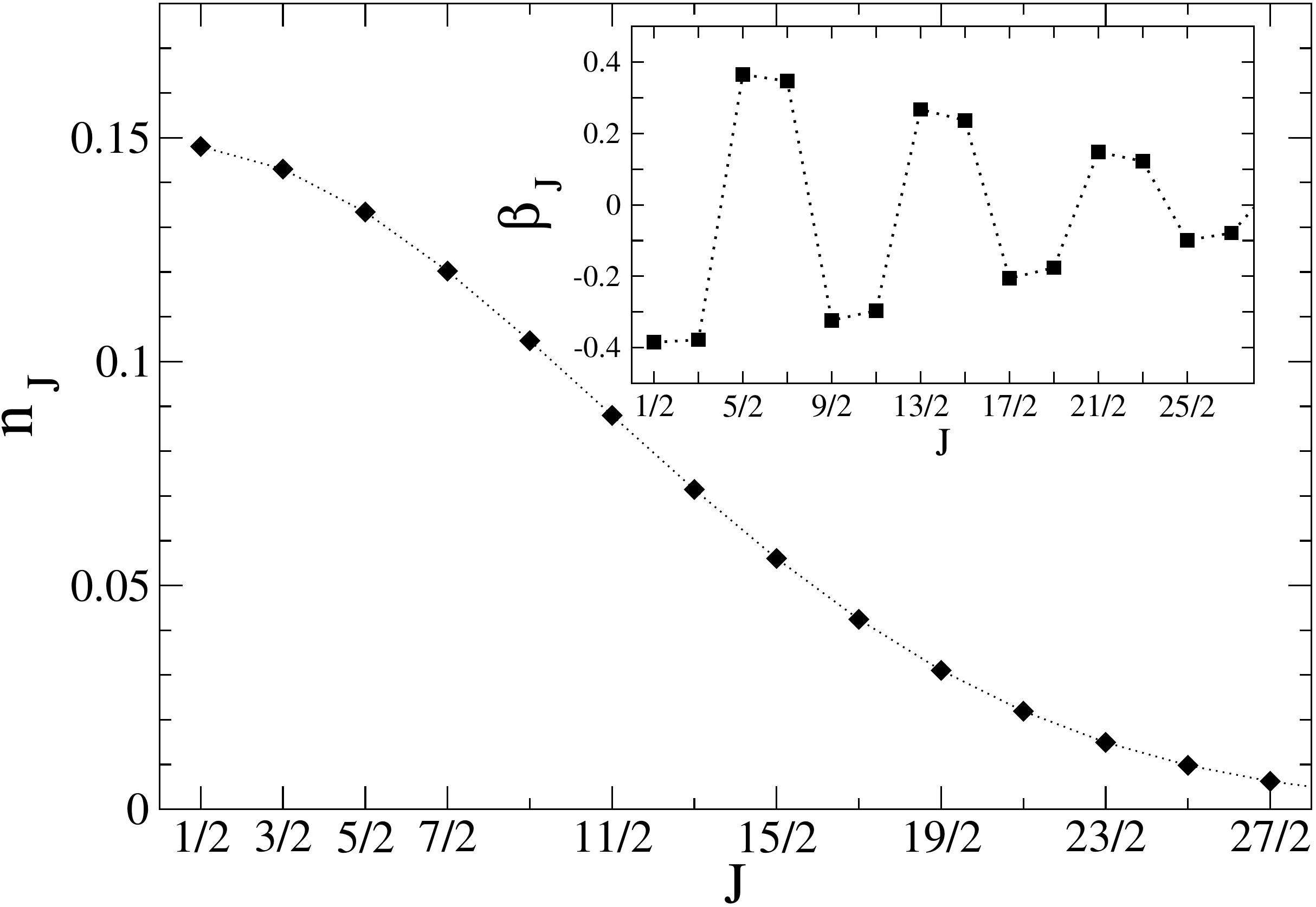}
\end{center}
\caption{\label{fig4} Distribution function, $n_J=\left|\beta^{}_J\right|^2$, 
vs $J$ for the $N=2$ ground state of $H_\infty$.  
We here take $\lambda\alpha^2=10^4$, which leads to 
${\cal E}_{\rm min}\simeq -0.0992725$.  
We stress that both $\beta_J$ (shown in the inset) as well as $n_J$ 
are \textit{independent}\ of the total angular momentum, $M_s=2M\hbar$.  
Dotted lines serve as guides to the eye only.  }
\end{figure}

For given total angular momentum, $M_s=2M\hbar$, 
we found in Sec.~\ref{sec3a} that
the eigenstate of $H_\infty$ with lowest energy can be constructed 
from the \textit{Ansatz} 
\begin{equation}\label{groundM}
|M\rangle= \sum_{J>0}  \beta_J \ c_{J+M}^\dagger 
c_{-J+M}^\dagger |0\rangle,
\end{equation}
with $\langle M'|M\rangle=\delta_{MM'}$.  Clearly, the
unmagnetized $M=0$ state in Eq.~\eqref{groundM} 
describes a superposition of time-reversed states, 
and thus preserves the TRS of the Hamiltonian. However, 
TRS is violated by all other states, $|M\ne 0\rangle$. 

Some algebra yields from Eq.~\eqref{tildebeta} the matrix elements
\begin{equation}\label{eig1}
\langle M'|H_\infty|M\rangle= 
\left( \frac{M^2}{\alpha^2}+ 2\lambda{\cal E}\right) \delta_{MM'} \hbar 
\omega,
\end{equation}
with the dimensionless ``energy''
\begin{equation}\label{eigne}
{\cal E} = \sum_{J,J'>0} \left( \frac{J^2}{2\lambda\alpha^2} \delta_{JJ'} +
 S_{J-J'}-S_{J+J'} \right)\beta_{J}^{} \beta_{J'}^{}.
\end{equation}
Since the matrix appearing in Eq.~\eqref{eigne} is real symmetric,
we can choose real-valued $\beta^{}_J$. 
Moreover, since the matrix is independent of $M$,
its lowest eigenvalue, ${\cal E}_{\rm min}$, is also $M$-independent and
depends on the interaction strength and on the Rashba coupling  
only through the combination $\lambda \alpha^2$. 
The corresponding normalized eigenvector is easily obtained numerically
and directly gives the $\beta_J^{}$.
Thereby we also obtain the normalized ground-state 
distribution function, $n_J=\left| \beta^{}_J\right|^2$.
Typical results for $\beta_J^{}$ and $n_J$ are shown in Fig.~\ref{fig4}.
We find a rather broad distribution function $n_J$,
very different from a Fermi function.
To reasonable approximation, the numerical results can be 
fitted to a Gaussian decay, $n_J\sim e^{-(J/J^*)^2}$, with $J^*\sim 
\sqrt{\alpha}$.  Since $J^*\ll \alpha$,
the relevant angular momentum states have $|J|\ll \alpha$, and
the single-band approximation is self-consistently fulfilled.
As shown in the inset of Fig.~\ref{fig4},
the $\beta_J$ exhibit a pairwise oscillatory behavior, 
where $\beta_J<0$ for $J=1/2$ and $3/2$, but $\beta_J>0$ for $J=5/2$
and $7/2$, and so on. 

\subsection{Ground state magnetization}\label{sec3c}

The above results indicate that for $\alpha\to \infty$ and given $M$, the
lowest energy is
\begin{equation}\label{einfm}
E^{(\infty)}_{M} = \left( \frac{M^2}{\alpha^2} 
+2\lambda {\cal E}_{\rm min} \right)\hbar\omega .
\end{equation}
While this suggests that the ground state has $M=0$, 
the $M^2/\alpha^2$ term (due to $H_0$) is in fact
subleading to Coulomb corrections beyond $H_\infty$, 
which approximately scale $\sim 1/\alpha$, see Eq.~\eqref{vappe}.
We therefore have to take these Coulomb matrix elements into account when 
determining the ground state.
To that end, using the symmetry relation (\ref{srel1}), and exploiting 
that $\hat M_s$ is conserved, we note that
$H_I$ [Eq.~\eqref{hintfin}] has the matrix elements
\begin{eqnarray}\label{hintmm}
&& \langle M'|H_I|M\rangle= 2 \delta_{MM'} \sum_{J,J'>0} 
\beta_J^{} \beta_{J'}^{} \\ \nonumber && \times
 \left( V^{(J-J')}_{-J+M,J+M}- V^{(J+J')}_{-J+M,J+M} \right).
\end{eqnarray}
Therefore, the energies $E_M^{(\infty)}$ in Eq.~\eqref{einfm} will
be independently shifted by this perturbation, and
the $(J_1,J_2)$-dependence of the Coulomb matrix elements becomes 
important, see Sec.~\ref{sec2d}. In particular,
terms with odd angular momentum exchange $m$ will contribute.  
Treating the Coulomb corrections in perturbation theory, 
the lowest energy for fixed $M$ is
\begin{equation}\label{emm}
E_M = E_{-M}= E^{(\infty)}_M +  
\langle M|H_I|M\rangle - 2\lambda {\cal E}_{\rm min} \hbar\omega,
\end{equation}
where $E_M=E_{-M}$ follows from the symmetry relations in Sec.~\ref{sec2d}.  
Comparing $E_M$ to the respective $M=0$ value, $\delta E_M=E_M-E_{M=0}$, 
we finally obtain 
\begin{equation}\label{dem}
\delta E_M = \frac{M^2}{\alpha^2} \hbar\omega + \langle M|H_I|M\rangle-
\langle M=0|H_I|M=0\rangle, 
\end{equation}
with $\langle M|H_I|M\rangle$ in Eq.~\eqref{hintmm}.

\begin{figure}
\begin{center}
\includegraphics[width=0.49\textwidth]{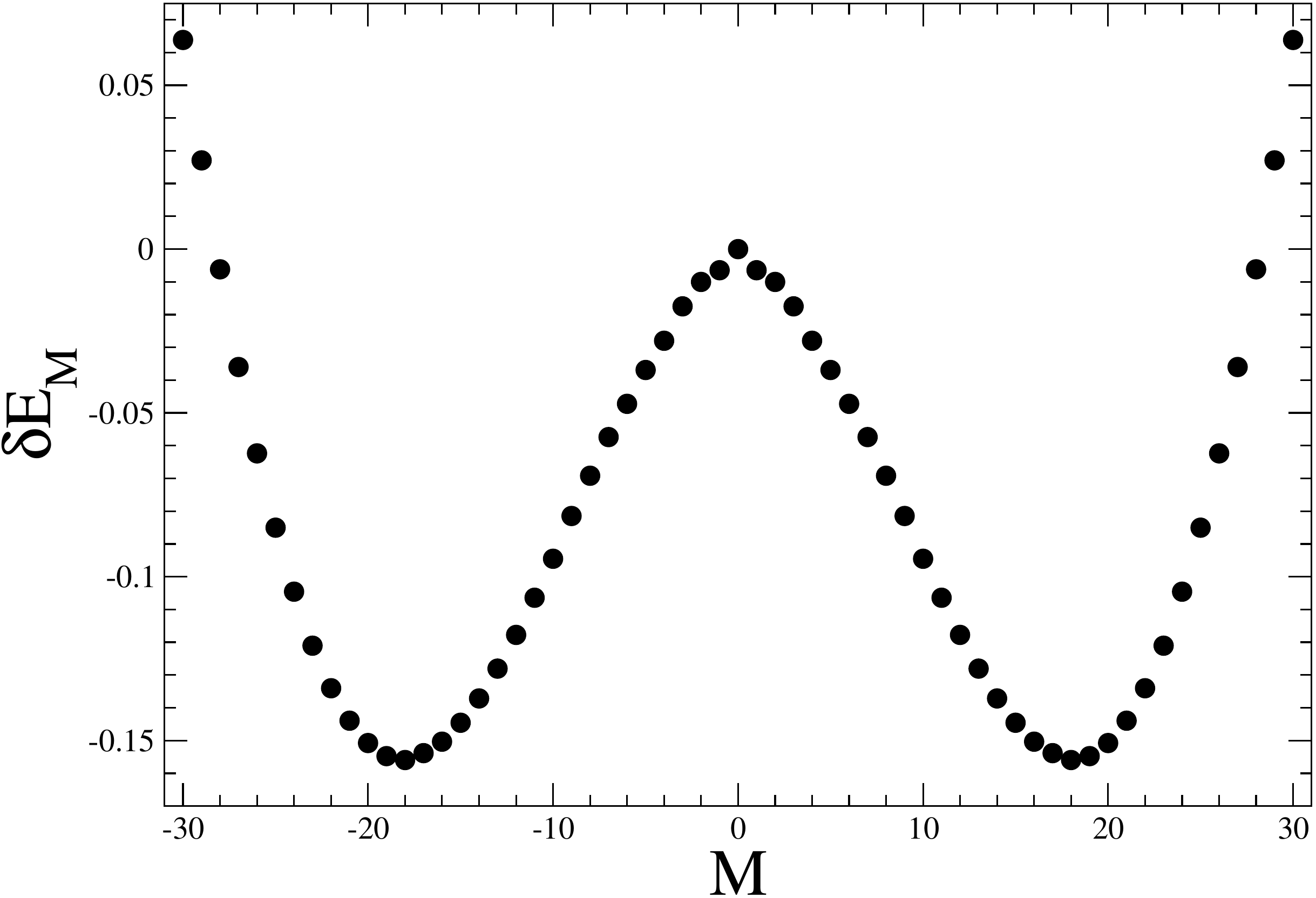}
\end{center}
\caption{\label{fig5}  Lowest two-particle energy for fixed $M$ 
relative to the $M=0$ state, $\delta E_M$ [Eq.~\eqref{dem}], vs $M$.   
We here consider $\alpha=30$ and $\lambda=1$, where $\delta E_M$ is given in
units of $\hbar\omega$.}
\end{figure}

The numerical result for $\delta E_M$ is shown
in Fig.~\ref{fig5}, where we take $\alpha=30$ and $\lambda=1$ as
concrete example.  Interestingly, the unmagnetized $M=0$ state
represents a local energy maximum.   This can be rationalized by noting
that the $\beta_J^{}$ have pairwise alternating sign, see Fig.~\ref{fig4}.  
For small but non-zero $|M|$, Eq.~\eqref{hintmm} is thus dominated by 
the $J=J'$ contribution due to $V^{(J-J')}_{-J+M,J+M}$, resulting
in the estimate
\begin{equation}\label{demap}
\delta E_M \approx 2\sum_{J>0} n_J \left( V^{(0)}_{-J+M,J+M} - 
V^{(0)}_{-J,J} \right) < 0 .
\end{equation} 
The inequality here follows by noting that Coulomb matrix elements 
with $m=0$ are always positive and have a maximum for $|J_2|=|J_1|$,
cf.~Sec.~\ref{sec2d}.  For large $|M|$, however,
the $M^2$ contribution in Eq.~\eqref{dem} becomes crucial.
As a consequence, we arrive at a symmetric double-well behavior for
$\delta E_M$, with two minima at $M=\pm M_0$.
This simple argument is consistent with our numerical 
results based on Eq.~\eqref{emm}, see Fig.~\ref{fig5}.

Since we typically find $M_0\gg 1$, see Fig.~\ref{fig5}, 
this effect must come from the orbital angular momentum.
The value of $M_0=M_0(\alpha,\lambda)$ can be estimated analytically 
as follows.  Evaluating $\delta E_M$ from Eq.~\eqref{dem}, employing
the approximation in Eq.~\eqref{demap} and the expression for
the matrix elements in Eq.~\eqref{vappe},  we find
\begin{equation}
\frac{\delta E_M}{ \hbar \omega} \simeq 
\frac{M^2}{\alpha^2} - \frac{2 \lambda}{\pi \alpha} 
\int_0^{\pi/2} d\varphi \frac{\sin \varphi}{\cos^2 \varphi} \sin^2(2M \varphi).
\end{equation}
The minimum of $\delta E_M$ with $M=+M_0$ then follows from the equation
\begin{equation}
M_0 = \frac{2 \lambda \alpha}{\pi} \int_0^{\pi/2} d\varphi
\frac{\varphi \sin \varphi}{\cos^2 \varphi} \sin(4 M_0 \varphi).
\end{equation}
Assuming $M_0 \gg 1$, the main contribution to the integral comes from
$\varphi \alt 1/M_0$, and performing the subsequent integration
implies $M_0 \approx (\lambda \alpha)^{1/4}$.  Clearly,
this suggests that $M_0$ can be very large even for weak interactions.

The ground state of the $N=2$ dot, 
\begin{equation}\label{eqphi}
|\Phi\rangle= \sum_\pm c_\pm|\pm M_0\rangle, \quad \sum_\pm |c_\pm|^2=1,
\end{equation}
is spanned by the two degenerate states $|\pm M_0\rangle$,
 with magnetization $M_s=\pm 2M_0\hbar$, respectively.
Unless $c_+c_-=0$, we note that $|\Phi\rangle$ is not an eigenstate
of the conserved operator $\hat M_s$. However, the magnetization 
expectation value,  $\langle \Phi|\hat M_s|\Phi\rangle= 2(|c_+|^2-
|c_-|^2) M_0 \hbar$, is finite except when $|c_+|=|c_-|$.
This suggests that by application of a weak magnetic field 
perpendicular to the 2D plane, the magnetization can be locked
to one of the two minima, say, $M_s=+2M_0\hbar$.  
Adiabatically switching off the magnetic field, we then 
expect $|\Phi\rangle=|M_0\rangle$,
since there is an energy barrier to the $|-M_0\rangle$ state.  
Since the barrier is not infinite, we cannot exclude
that quantum-mechanical tunneling effects will ultimately 
establish an unmagnetized ground state with $|c_+|=|c_-|$,
in particular when taking into account violations of the
perfect rotational symmetry of the dot assumed in our model.
For instance, such imperfections correspond to an eccentricity of the 
confinement potential or the presence of nearby impurities.
As long as such imperfections represent a weak perturbation,
however, the associated tunneling timescales connecting
the two degenerate states $|\pm M_0\rangle$ are expected to be very long.
We shall discuss this issue in some detail in Sec.~\ref{sec5}.  

For practical purposes, assuming that quantum tunneling is not 
relevant on the timescales of experimental interest, 
adiabatically switching off the magnetic field then effectively results
in the ground state $|\Phi\rangle=|M_0\rangle$.
This state carries a large magnetization, $M_s=2M_0\hbar$, 
and thus also a circulating charge current.  
Such a state appears to spontaneously break the TRS of the 
Hamiltonian, and is interpreted here 
as a ``molecular'' \textit{orbital ferromagnet}.  

The above discussion pertains to the idealized $T=0$ case.
In practice, the zero-temperature limit also governs the physics at 
temperatures well below the above energy barrier, 
 $k_B T \ll  |\delta E_{M_0}|$.  However, at higher temperatures,
thermally induced transitions between both minima happen on short 
timescales, and the overall magnetization of the dot vanishes.  
Nonetheless, $\hat M_s^2$ still has a finite expectation value.

\subsection{Spin and charge density}

Before proceeding with a discussion of numerical results for $N>2$,
let us briefly address the spin and charge density 
for $\alpha\to \infty$. We  assume that the $N=2$ system is in a definite 
ground state, say $|M_0\rangle$. 

The total spin density at
position $\boldsymbol{r}=r(\cos\theta,\sin\theta)$ follows as
\begin{equation}\label{spindens}
\boldsymbol{S}(r,\theta) = \sum_{J>0} n_J \left[ \boldsymbol{s}_{J+M_0}
(r,\theta) + \boldsymbol{s}_{-J+M_0} (r,\theta)\right],
\end{equation}
where $\boldsymbol{s}_J=(s^x_J,s^y_J,s^z_J)$ is the spin density
for the single-particle state $\tilde \psi_J(r,\theta)$. Using 
Eq.~\eqref{wlar}, we obtain, e.g.,
\begin{equation}
s^x_J(r,\theta) \simeq \frac{\hbar}{2} 
\frac{e^{-r^2/l_T^2}}{\pi^{3/2} l_T r} \cos(\theta) 
\sin(2k_0 r-\pi J).
\end{equation}
As a consequence, the two contributions in Eq.~\eqref{spindens} precisely 
cancel each other, and $S^x=0$. By the same argument, we also find that
the $y$- and $z$-components of the spin density vanish.  
In the limit $\alpha\to \infty$, the spin density $\boldsymbol{S}$ 
is therefore identically zero.  In practice, 
finite contributions may come from subleading ($\sim 1/\alpha$) terms, 
but these are small for $\alpha\gg 1$. 

We now turn to the charge density, $\rho_c(r)$, which is 
always radially symmetric.  For $\alpha\to \infty$, all single-particle 
states, $\tilde\psi_J$ in Eq.~\eqref{wlar}, lead to the
 \textit{same} probability density in space, and we therefore conclude that 
$\rho_c(r)$ must be independent of Coulomb interactions.  For $\lambda\alt 1$ 
and arbitrary particle number $N$, we thus obtain
\begin{equation}\label{chargedens}
\rho_c(r)= \frac{eN}{\pi^{3/2}l_T r} e^{-r^2/l_T^2} ,
\end{equation}
which satisfies the expected normalization, 
$2\pi\int_0^\infty dr r \rho_c(r)=eN$.
We mention in passing that the ``edge'' state property of the single-particle
states, i.e., states with larger $|J|$ live further away from
the dot center, can be seen from the finite-$\alpha$ wavefunctions 
in Eq.~\eqref{Fm} \cite{li1}, but not anymore 
from their asymptotic $\alpha\to \infty$ 
form in Eq.~\eqref{wlar}.
The $\lambda$-independence of $\rho_c(r)$ at large $\alpha$ is in
marked contrast to the case of weak spin-orbit coupling, where $\rho_c$
contains information about interactions and 
can be used to detect Wigner molecule  formation
\cite{egger,weissegger}.  Instead, the charge density in Eq.~\eqref{chargedens} 
is featureless for arbitrary $N$, pointing once again
to the absence of the Wigner molecule  for $\alpha\gg 1$ and $\lambda\alt 1$.  
Finally, we note that by 
computing the pair distribution function \cite{giul}, 
we also find no trace of Wigner molecule formation in this limit.

\section{Exact diagonalization and Hartree-Fock calculations}
\label{sec4}

We now discuss numerical results for the ground-state energy
and magnetization for $N\le 10$ electrons in the dot.
These results were obtained by means of the standard exact diagonalization 
technique and by Hartree-Fock (HF) theory from $H=H_0+H_I$,
with $H_0$ in Eq.~\eqref{h02q},  $H_I$ in Eq.~\eqref{hintfin}, and
the Coulomb matrix elements \eqref{Vmm},
see Sec.~\ref{sec2}. This implies that the following results are  not 
restricted to the $\alpha\to \infty$ limit considered in Sec.~\ref{sec3}. 
However, we are limited to rather weak interactions, $\lambda\alt 1$, 
and moderate particle numbers, $N<\alpha$, because of our single-band
approximation.  We first describe our exact diagonalization results 
and then turn to HF theory.

\subsection{Exact diagonalization}\label{sec4a}

\begin{figure}
\begin{center}
\includegraphics[width=0.5\textwidth]{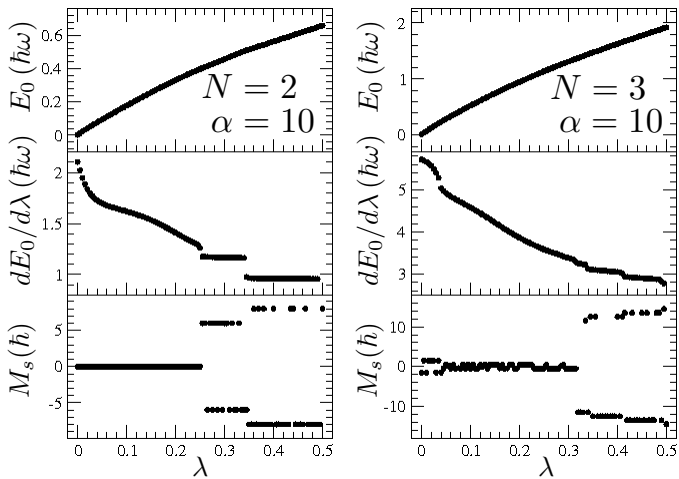}
\end{center}
\caption{\label{fig6} Exact diagonalization results for the
 ground-state energy, $E_0$, and the magnetization, $M_s$, 
vs interaction parameter $\lambda$, for 
$\alpha=10$ and $N=2$ (left) and $N=3$ (right).  
The top row shows $E_0(\lambda)$, 
where singular structures are visible in the derivative, $dE_0/d\lambda$, as
depicted in the center panel. The bottom panel shows the 
magnetization, $M_s$,  where  $M_s\ne 0$ for $\lambda>\lambda_c$.  
(For $N=3$, a finite but very small value for 
$M_s$ numerically appears already for $\lambda_c < 0.31$.) }
\end{figure}
  
Using the Rashba parameter $\alpha=10$, exact diagonalization
results for $N=2$ and $N=3$ electrons in the dot 
are shown in Fig.~\ref{fig6}.  While $E_0(\lambda)$
at first sight seems rather featureless (top panel), there are non-analytic
features that become visible when plotting the first derivative 
(center panel).  Let us discuss this point in detail for $N=2$, see the 
left side of Fig.~\ref{fig6}.
The first non-analytic feature occurs at $\lambda_c\approx 0.25$, 
where the second derivative diverges,
$d^2E_0/d\lambda^2\to -\infty$, as the interaction parameter $\lambda$ 
approaches the critical value $\lambda_c$ from below.  
In close analogy to the results obtained from $H_\infty$ 
in Sec.~\ref{sec3}, the ground state for $\lambda>\lambda_c$ has
the magnetization $M_s= \pm 2M_0\hbar$, where $M_0$ is integer
and the ground state is degenerate with respect to both signs.
In the exact diagonalization, the ``initial conditions''
selecting the eventually realized state ($M_s=+2M_0\hbar$ 
or $M_s=-2M_0\hbar$) correspond to unavoidable numerical rounding errors.
In contrast to the $\alpha\to\infty$ limit, however, the 
interaction parameter $\lambda$ must now exceed a critical value, 
$\lambda_c$, to allow for the orbital ferromagnet.
For $\lambda<\lambda_c$, the $M=0$ state is the ground state, which
is adiabatically connected to the noninteracting ground state.  
Since energy levels of states with different conserved total angular
momentum $M_s$ can cross each other, 
the critical value $\lambda_c$ marks a quantum phase transition.
However, once disorder or eccentricity of the quantum dot are present,
angular momentum conservation breaks down and the transition will
correspond to a smooth crossover phenomenon.
The observed large value of the magnetization, $|M_s|=6\hbar$ 
for $\lambda>\lambda_c$,  see Fig.~\ref{fig6}, again rules out a 
purely spin-based explanation. In fact, additional jumps 
to even higher $|M_s|$ are observed for larger $\lambda$ in Fig.~\ref{fig6}.

Similar features are also observed for $N=3$, where
exact diagonalization results are shown on the right
side of Fig.~\ref{fig6}.
Again, the first derivative of $E_0(\lambda)$ displays non-analytic
behavior. For small $\lambda$, the state stays close to a
doubly degenerate Fermi sea, see Sec.~\ref{sec2a}.
For $\lambda>\lambda_c\approx 0.31$, however, a large  
magnetization emerges, $|M_s|= 11.5 \hbar$. 

\begin{figure}
\begin{center}
\includegraphics[width=0.5\textwidth]{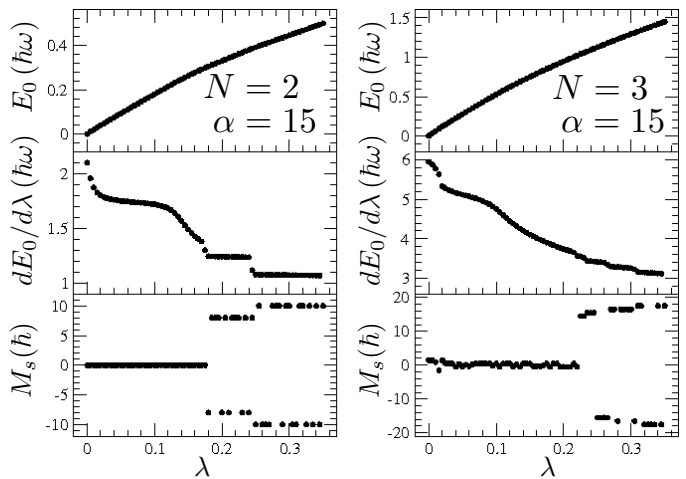}
\end{center}
\caption{\label{fig7} Same as Fig.~\ref{fig6} but for $\alpha=15$.}
\end{figure}

The results of Sec.~\ref{sec3} show that the critical interaction strength
$\lambda_c$ vanishes for $\alpha\to \infty$.
We therefore expect $\lambda_c$ to decrease with increasing $\alpha$.
To study this point, exact diagonalization results
 for $\alpha=15$ are shown in Fig.~\ref{fig7}.
All qualitative features observed for $\alpha=10$ are recovered, and the 
critical value $\lambda_c$ is indeed found to decrease:  For 
$N=2$, we now find $\lambda_c\approx 0.17$ (instead of $\lambda_c\approx 
0.25$ for $\alpha=10$), while for $N=3$, we obtain $\lambda_c\approx
0.22$ (instead of $\lambda_c\approx 0.31$).  This confirms
that with increasing spin-orbit coupling strength,
 the orbital ferromagnetic state is reached already for weaker interactions.

\subsection{Hartree-Fock calculations}\label{sec4b}

Finally, let us turn to numerical results for larger $N$, where 
exact diagonalization becomes computationally too expensive.
We have carried out an unrestricted Hartree-Fock analysis following
the textbook formulation \cite{giul}, in order to find 
the energy and the total angular momentum of the $N$-electron ground state. 
We note that HF calculations are known to provide a reasonable
description for $\alpha=0$ \cite{landman,reusch1,reusch2}.  
We find that a diagonal HF density matrix is sufficient, with 
variational parameters $n_J = \langle c_J^\dagger c_J^{}\rangle$ for
half-integer $J$ subject to the normalization condition $\sum_J n_J=N$.
Up to a constant, the resulting HF Hamiltonian is given by
\begin{equation}
H_{\rm HF} = \sum_J \left(  E_J+   2\sum_{J'} 
\left [ V_{J,J'}^{(0)}-V_{J,J'}^{(J'-J)} \right] n_{J'}^{} \right) c_J^\dagger
c_J^{}.
\end{equation}
The self-consistent HF ground state is numerically found 
by iteration, starting from randomly chosen initial distributions.
The converged $\{ n_J\}$ distribution yields the magnetization, $M_s$, and
the ground-state energy. 
For $\lambda$ approaching
the (HF value of the) critical interaction parameter, $\lambda_c(\alpha,N)$,
the energy shows similar non-analytic features as found
from exact diagonalization, see Sec.~\ref{sec4a}.
For $\lambda>\lambda_c$, a large ground-state magnetization 
is observed, again corresponding to orbital ferromagnetism.

\begin{figure}
\begin{center}
\includegraphics[width=0.45\textwidth]{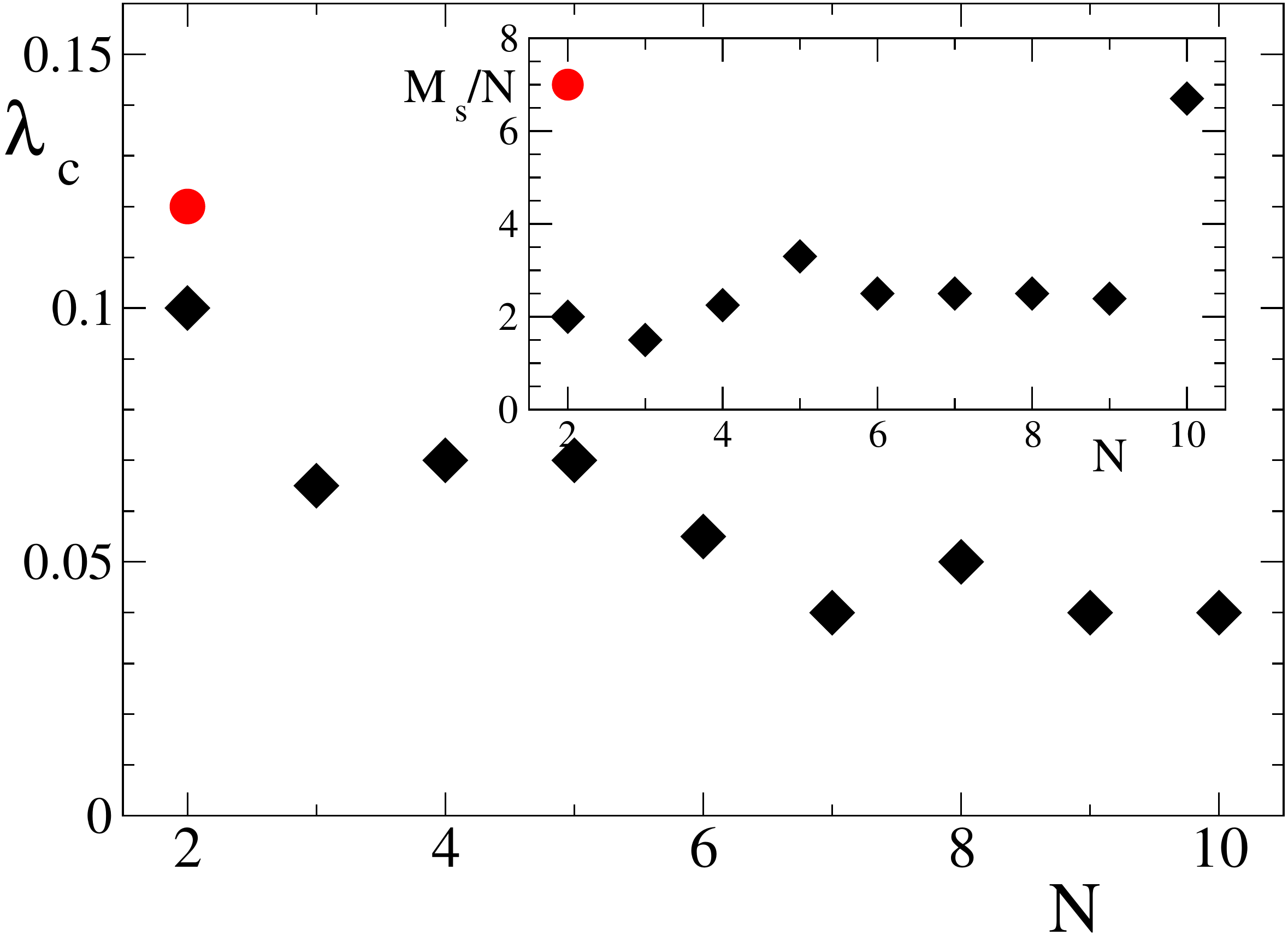}
\end{center}
\caption{\label{fig8} Hartree-Fock results (black diamonds) for 
the critical interaction strength $\lambda_c$ vs particle number $N$
for $\alpha=30$ (main panel). The red circle shows the 
corresponding exact diagonalization result for $N=2$.
Inset: Magnetization $M_s$ (in units of $\hbar N$) found 
for $\lambda\agt \lambda_c$, vs particle number $N$. 
}
\end{figure}
  
Our HF results for $\lambda_c$ and $M_s$ are shown in Fig.~\ref{fig8}.
We consider the Rashba spin-orbit coupling $\alpha=30$, and
up to $N=10$ electrons in the dot.  Unfortunately,
we cannot address larger $N$, for otherwise our 
single-band approximation is not justified anymore.  
For $N=2$, the corresponding
exact diagonalization values are also given.  The HF value
for $\lambda_c$ is only slightly smaller than the exact one,
which suggests that HF theory is at least qualitatively useful.
That the HF prediction is below the exact one for $N=2$ can be rationalized
by noting that HF theory generally tends to favor ordered phases such as
orbital ferromagnetism, resulting in a smaller value for $\lambda_c$.
The magnetization for $\lambda>\lambda_c$, however, is 
a more difficult quantity to predict due to the shallow minima
of the free energy curves in Fig.~\ref{fig5}.  Indeed, the inset
of Fig.~\ref{fig8} shows that the HF value of the magnetization (which appears
to scale as $M_s\sim N$) is significantly smaller than the exact one.
With increasing $N$, the HF predictions for $\lambda_c$ indicate
that the transition to the orbital ferromagnet persists. Moreover, 
this transition can even be reached at weaker interactions.

\section{Discussion}\label{sec5}

In this work, we have studied the interacting $N$-electron problem 
for a parabolic 2D quantum dot (with $N\le 10$) in the limit of strong
Rashba spin-orbit coupling, $\alpha\gg 1$.  
This regime is characterized by an almost flat single-particle
spectrum, where we find that already weak-to-intermediate Coulomb interactions
(our single-band approximation permits us to study the regime $\lambda\alt 1$
only) are sufficient to induce molecular orbital ferromagnetism.  
This state is observed for $\lambda>\lambda_c( \alpha, N)$,
where our $N=2$ solution in Sec.~\ref{sec2} shows that
the critical strength $\lambda_c\to 0$ for $\alpha\to \infty$. 
For finite (but large) $\alpha$, however, $\lambda_c$ will be finite.  
The orbital ferromagnet has a giant total angular momentum, 
accompanied by a circulating charge current. 

Coming back to our discussion in Sec.~\ref{sec3c}, we now address 
issues concerning the experimental observation of the predicted
orbital ferromagnetism for a single quantum dot.  
The transition to this state could be induced in practice by 
varying the electrostatic confinement potential and/or the 
gate-controlled Rashba spin-orbit coupling in order to reach the
regime defined by $\alpha\gg 1$ and $\lambda>\lambda_c$.  
By allowing for an eccentricity of the dot confinement
potential, which also can be achieved with appropriate gate voltages, 
quantum tunneling processes connecting the free energy minima with 
opposite magnetization, $M_s=\pm M_{\rm min}$, are expected to 
become relevant, see Sec.~\ref{sec3c}.  
The corresponding timescale for such processes can be 
estimated as follows.  We first note that the free energy barrier 
between both minima, $B\hbar \omega$, corresponds to a number 
$B\approx 0.1\ldots 0.15$, see Fig.~\ref{fig5}.  
We next employ a paradigmatic effective low-energy model to include the effects
of imperfections breaking the ideal rotational symmetry,
\begin{equation}\label{heffeq}
H_{\rm eff}=  \left[ \frac{\epsilon}{2} \phi^2 + B
\frac{(M_s^2- M_{\rm min}^2)^2}{M_{\rm min}^4} \right] \hbar\omega.
\end{equation}
The first term describes the dot eccentricity, with   
a small dimensionless parameter $\epsilon$, where the polar angle $\phi$ 
is conjugate to the magnetization $M_s$. 
The second term approximates the double-well potential in Fig.~\ref{fig5}.
The two lowest eigenenergies for $H_{\rm eff}$ are 
known exactly \cite{alexal}.  From the result, we find the level splitting
\begin{equation} 
\delta E = \sqrt{\frac{2}{\pi}} 64 B\hbar\omega 
\exp\left(-\frac{4\sqrt{2B}}{3\sqrt{\epsilon}} M_{\rm min} \right) .
\end{equation}
The resulting timescale for tunneling processes, $\tau$, 
is thereby estimated as 
\begin{equation}
\omega\tau= \frac{\hbar\omega}{\delta E} \approx 0.2 
e^{ 5.96  M_{\rm min} }.
\end{equation}
where, for simplicity, we have put $B=0.1$ and $\epsilon=0.01$.
For the value $M_{\rm min}\approx 18$ observed in Fig.~\ref{fig5}, we get 
the estimate $\omega\tau\approx 10^{45}$.  This astronomically long 
tunneling time strongly suggests that on experimentally 
accessible timescales, the orbital ferromagnet described in this paper
will be indistinguishable from a true equilibrium state.

It is also useful to contrast the behavior reported here to the well-known 
persistent currents in normal-metal quantum rings 
\cite{buttiker,gogolinring,tapashring,hauslerring,tanring}, 
where a circulating equilibrium electric current flows 
and can be experimentally detected, see Ref.~\cite{pcexp} and references
therein.  First, a persistent current flows
already in noninteracting quantum rings but requires a nonzero flux
threading the ring, while orbital ferromagnetism in a 2D dot is 
generated by the interplay of Coulomb interactions and 
strong spin-orbit coupling. 
Second, the total angular momentum (magnetization) predicted here
for a 2D Rashba dot can be very large.  Therefore, 
the circulating currents in our case should exceed the persistent currents
observed in quantum rings by far.  Despite of these differences, 
the persistent current analogy 
also suggests ways to observe our predictions experimentally.  Another 
possibility is to study the response to a weak magnetic field applied 
perpendicular to the 2D plane.  The low-field susceptibility is then 
expected to be singular, just as in an ordinary ferromagnet.
At elevated temperatures approaching the free energy barrier height 
discussed above, the orbital magnetization in our system will be 
thermally suppressed and ultimately disappear.  
The relevant temperature scale for this
crossover is $T_c\approx B\hbar\omega/k_B$. For typical quantum dots 
\cite{kouwenhoven}, $T_c\approx 1$ to 10~K. 

To conclude, we hope that our prediction of orbital ferromagnetism in 
Rashba dots will stimulate further theoretical and experimental
work.  For instance, it remains an open question to address the
transition from the orbital ferromagnet to a Wigner molecule with 
increasing interaction strength for large Rashba coupling.  In order
to achieve this description, one needs to go beyond the single-band
approximation employed in this work.

\acknowledgments
We thank W.~H\"ausler for discussions.
This work has been supported within the networks SPP 1666 and SFB-TR 12 of
the Deutsche Forschungsgemeinschaft (DFG).

\appendix

\section{On ultra-strong Rashba couplings}

In this Appendix, we address an alternative calculation of the interaction 
matrix elements $V_{J_1,J_2}^{(m)}$ for $\alpha\to \infty$.
Instead of taking this limit as $k_0\to \infty$ with finite
$l_T$, see Sec.~\ref{sec2c}, we here formally assume 
a fixed spin-orbit momentum $k_0$ but large $l_T$.  
The $\alpha\to \infty$ limit taken in this manner is subtle since (i) the 
resulting expressions require infrared 
regularization with $l_T$ setting the effective system size, and (ii) the 
single-band approximation requires a finite confinement frequency 
in order to be justified. However, it is also beneficial 
since one can proceed directly in momentum space and thereby obtain an
intuitive understanding of the parity effect.  

We start by noting that the states \eqref{psiJ} describe a 
Gaussian distribution of the probability density in momentum space 
around $k=k_0$, where $1/l_T$ sets 
the amplitude of zero-point fluctuations of $k$ around $k_0$.  
For $l_T\to \infty$, this density becomes
$\left| \psi_J({\bf k}) \right|^2 \simeq (2 \pi/k_0)\ \delta (k - k_0)$.
The states \eqref{psiJ} thus have the limiting  behavior
\begin{equation}
\psi_J({\bf k}) \simeq \sqrt{\frac{2\pi^{3/2}}{k_0 l_T}}
e^{i (J-1/2)\phi} \delta (k - k_0) \left( \begin{array}{c} 1 \\ 
-i e^{i \phi} \end{array} \right) ,
\end{equation}
describing localization on a ring in momentum space.
The interaction Hamiltonian is
\begin{equation}
H_I = \frac12  \int \frac{d^2 {\bf q}}{ (2 \pi)^2} \, 
\frac{2\pi e^2}{\varepsilon_0 q} : \rho(-{\bf q}) \rho({\bf q}):,
\end{equation}
where $::$ denotes normal ordering and   
\begin{equation}
\rho({\bf q})= \int\frac{d^2 {\bf k}}{(2 \pi)^2} \,
\Psi_{{\bf k} + {\bf q}}^\dagger \Psi^{}_{{\bf k}},\quad
\Psi_{\bf k}= \sum_J \psi_J({\bf k}) c_J.
\end{equation}
Writing  
\begin{equation}
{\bf k} = k \left (\begin{array}{c}\cos \phi\\ \sin \phi\end{array}\right),
\quad
{\bf q} = q \left(\begin{array}{c} \cos \vartheta\\ \sin \vartheta\end{array}
\right), \quad {\bf k}'= {\bf k}+{\bf q}, 
\end{equation}
it is now crucial to take into account 
the constraints $k=k^\prime=k_0$ coming from the 
$\delta$-functions in $\psi_J({\bf k})$. In effect, all momenta for
 incoming, ${\bf k}_{1,2}$, and outgoing, ${\bf k}^\prime_{1,2}$, 
electrons must be located on a ring of radius $k_0$ in momentum space. 
This severe phase-space restriction is only met by two types of 
interaction processes as explained next. 

\begin{figure}[h]
\begin{center}
\includegraphics[width=0.55\textwidth]{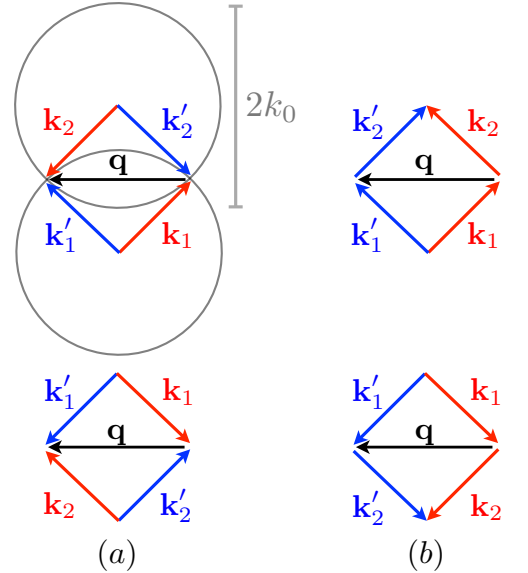}
\end{center}
\caption{\label{figapp}  Coulomb two-particle scattering processes, 
${\bf k}_{1,2} \to {\bf k}_{1,2}^\prime = {\bf k}_{1,2} \pm {\bf q}$,
for given exchanged momentum, ${\bf q}$, in the $\alpha\to\infty$ limit, 
where all particle momenta are constrained to a ring of radius $k_0$.  
For ${\bf q}\ne 0$, only four interaction processes are possible,
which can be grouped into two classes:  (a)
BCS-like scattering of a pair of opposite-momentum states, 
${\bf k}_1 = - {\bf k}_2$, into another pair with 
${\bf k}_1^\prime = - {\bf k}_2^\prime$,
and (b) exchange-type scattering processes,  
$({\bf k}_1,{\bf k}_2)\to({\bf k}_1^\prime = 
{\bf k}_2,{\bf k}_2^\prime = {\bf k}_1)$.}
\end{figure}

Shifting the integration variable 
$\phi\to \phi+\vartheta$, some algebra yields (integer $m$) \cite{mathfoot}
\begin{eqnarray}\label{eqvv}
&& V_{J_1,J_2}^{(m)} = \frac{e^2 }{4\varepsilon_0 l_T^2} 
\int_0^{2 k_0} d q \int_0^{2 \pi} 
\frac{d \phi_1 d \phi_2}{2 \pi} \\ \nonumber && \times~~ 
\delta (k_1^\prime - k_0) \ \delta (k_2^\prime - k_0)
e^{im(\phi_2-  \phi_1)} \sum_{\sigma_1, \sigma_2 = \pm} \\  &&\times~~
\left( 1 + \frac{q}{k_0} e^{i \phi_1} \right)^{J_1+m+\sigma_1/2}
\left( 1 - \frac{q}{k_0} e^{i \phi_2} \right)^{J_2-m+\sigma_2/2},
\nonumber
\end{eqnarray}
where the $\delta$-function implies the constraint
\begin{equation}
k_1^\prime (\phi_1) = \sqrt{k_0^2 + q^2 + 2 k_0 q \cos \phi_1} = k_0,
\end{equation}
and similarly for $k_2'$.  This leads to the condition
$\cos \phi_1 = - \cos \phi_2 = - q/2k_0$,
which is met by two types of scattering processes only, namely 
(a) for $\phi_2 = \pi + \phi_1$ (BCS-like pairing), and 
(b) for $\phi_2 = \pi - \phi_1$ (exchange-type process),
see Fig.~\ref{figapp}. Such spin-orbit-induced constraints on 
interaction processes were also recently pointed out in Ref.~\cite{scring}.
Parametrizing $q=2k_0\cos\varphi$ in Eq.~\eqref{eqvv}, we obtain 
\begin{eqnarray}\label{vappe}
&& V_{J_1,J_2}^{(m)} = (-1)^{J_1 + J_2+m-1}
\frac{e^2}{2 \pi\varepsilon_0 k_0 l^2_T}
\int_{0}^{\pi/2} d\varphi  
 \\ \nonumber &&\times~~ \sin\varphi \frac{- 
\cos [2 ( J_1 + J_2) \varphi] +
\cos  [2( J_1 - J_2 + m) \varphi ]}{\cos^2\varphi} ,
\end{eqnarray}
where the first (second) term in the numerator results from 
BCS-like (exchange-type) processes. Importantly,  
the above integral is infrared divergent for
$q=2k_0\cos\varphi \to 0$. To regularize this singularity, we 
employ $l_T$ as effective system size and require $ql_T>1$.
After some algebra, we find the ($J_1,J_2$)-independent result
$V_{J_1,J_2}^{(m)} \simeq \lambda\hbar\omega \delta_{m,\textrm{even}}$,
which recovers the parity effect in Sec.~\ref{sec2c}, 
including the $(J_1,J_2)$-independence of the matrix elements. 
In contrast to Eq.~\eqref{vmla}, however, the even-$m$ 
Coulomb matrix elements found here are also independent of $m$.
This indicates that the limits $k_0\to \infty$ and 
$l_T\to \infty$ do not commute.


\begin{thebibliography}{12}
\bibitem{footnew}
The angular part of this state represents the
low-energy eigenspinor of an effective Hamiltonian,
$H_{\rm eff}=-\partial^{2}_{\phi}-\Delta \mathcal{P}_{h}$
(in arbitrary energy units) with $|\Delta| \gg 1$, which
describes a particle on a ring in momentum space with Rashba 
spin-orbit coupling.
\bibitem{foot1}
The direct numerical integration of Eq.~(\ref{Vm1m2m}),
without performing the $\theta$-integration leading to Eq.~\eqref{Vmm},
is unstable because of the highly oscillatory integrand.
Fortunately, Eq.~\eqref{Vmm} does not suffer from such a problem.
\bibitem{footsm}
Empirically, numerical results for $S_m$ with $m\ne 0$
are well approximated by $S_{m} \simeq (2S_2/|m|) \delta_{m,{\rm even}}.$
\end{thebibliography}
\end{document}